\documentclass[a4paper,aps,prx,twocolumn,floatfix,showpacs,superscriptaddress,footnoteinbib]{revtex4-1}
\usepackage{graphics}
\usepackage{epsfig}
\usepackage{times}
\usepackage{xcolor}   
\usepackage{amsfonts}
\usepackage{amsmath}
\usepackage{amssymb}
\usepackage{amsthm}
\usepackage{hyperref} 
\usepackage{wasysym}
\usepackage{bm}

\begin{document}
\title{Branes and the Swampland}

\author{Hee-Cheol Kim}
\affiliation{Department of Physics, POSTECH, Pohang 790-784, Korea.}
\author{Gary Shiu}
\affiliation{ Department of Physics, 1150 University Avenue, University of Wisconsin, Madison, WI 53706, USA.}
\author{Cumrun Vafa}
\affiliation{Jefferson Physical Laboratory, Harvard University, Cambridge, MA 02138, USA.}

\begin{abstract}

Completeness of the spectrum of charged branes in a quantum theory of gravity naturally motivates the question of whether consistency of what lives on the branes can be used to explain some of the Swampland conditions.  In this paper we focus on consistency of what lives on string probes, to show some of the theories with ${\cal N}=(1,0)$ supersymmetry in 10d and 6d, which are otherwise consistent looking, belong to the Swampland. 
Gravitational and gauge group anomaly inflow on these probes can be used to compute the gravitational central charges  $(c_L,c_R)$ as well as the level of the group's current algebra $k_L$.  The fact that the left-moving central charge on the string probes should be large enough to allow {\it unitary} representations of the current algebra with a given level, can be used to rule out some theories.
 This in particular explains why it has not been possible to construct the corresponding  theories from string theory. 

\end{abstract}
\maketitle


\section{Introduction}
Increasing evidence points to the fact that some consistent looking theories cannot emerge as the IR limit of a quantum gravitational theory, and belong to the Swampland (see \cite{Brennan:2017rbf, Palti:2019pca} for a recent review for some of the Swampland criteria).  Ultimately we would like to explain {\it  why} the swampland conditions are necessary for consistency of quantum gravitational theories.  There are varying degrees of understanding for different swampland criteria.  In this paper we take a small step to initiate a new direction for a deeper understanding of the swampland criteria: we use consistency of brane probes to explain why certain consistent looking supergravity theories coupled to matter that were conjectured not to exist, indeed belong to the Swampland. See \cite{Banks:1997zs} (also \cite{Taylor:2011wt} for a discussion of its generalization) for an early idea of using string and brane probes to constrain Type I' string theory.

We focus on ${\cal N}=(1,0)$ supergravity theories in 10d and 6d (with 16 and 8 supercharges respectively). These theories enjoy the following common properties:  the gauge and gravitational anomaly cancellations severely limit the allowed possibilities.  In the 10d case we are limited to 4 choices for gauge groups \cite{Green:1987mn}: $E_8\times E_8, SO(32), E_8\times U(1)^{248},U(1)^{496}$. The latter two theories were conjectured to belong to the Swampland in \cite{Vafa:2005ui}.  An argument for this was presented in \cite{Adams:2010zy}.  Here we present an independent argument ruling out the latter two theories by showing that the left-moving central charge on the BPS strings in these theories, which should carry the current algebra of the corresponding group, is too small to realize the latter two theories.

Similarly anomaly cancellations for 6d (1,0) theories were used to show \cite{Kumar:2010ru} that there are rather restricted set of choices for the allowed gauge groups and matter representations.
Many of these were realized through F-theory.  But it was found that there are infinitely many examples that cancel anomalies but seem not to arise in F-theory or any other string realization. 
These  sets arose by having an unbounded rank for the gauge group or an unbounded number of tensors or choices of exotic representations.
In this paper we show that a subset of these theories that could not be realized in F-theory indeed belong to the Swampland.  In particular it was shown there \cite{Kumar:2010ru} that theories with $SU(N)\times SU(N)$ gauge group with two bifundamental matter representations and additional neutral matter are anomaly free for any $N$.  
However only $N\leq 8$ has been realized in string theory.  We show that indeed all the theories with $N>9$ belong to the Swampland by showing that the central charge of the $SU(N)\times SU(N)$ current algebra on certain BPS strings, which should exist due to the completeness assumption for the spectrum in a gravitational theory \cite{Polchinski:2003bq, Banks:2010zn} (see also \cite{Harlow:2018tng}), are too small to lead to unitary representations for these cases. Moreover it was found that a family of models with an unbounded number of tensors $T=8k+9$ and gauge group $(E_8)^k$, even though their anomalies cancel,  cannot be realized in F-theory except for $k<3$. We show that for a similar reason all these theories are ruled out.  

We view this work as just the beginning of the program of using brane probes for a deeper understanding of the swampland conditions. As a first step, we demonstrate the power of this approach with a few examples and with only string probes, but we expect this program has wider applicabilities in delineating the landscape from the swampland.   One can in principle consider not just the unitarity of the matter content on the branes, but also consistency between various types of branes and their interactions with one another as other possible ways to better understand the swampland conditions.

The organization of this paper is as follows.  In Section \ref{10d},
we discuss the consistency conditions of string probes for ${\cal N}=(1,0)$ supergravity theories in 10d,
and show that the two anomaly free theories with $E_8\times U(1)^{248}$ and $U(1)^{496}$ gauge groups are in the Swampland. In Section \ref{6d}, we discuss similar consistency conditions for 6d theories and show that unitarity of the 
current algebra on the string probes can be used to 
rule out several infinite families of anomaly free 6d ${\cal N}=(1,0)$ supergravity theories.
We conclude in Section \ref{conclusions}.
Some details are relegated to the appendices.

\section{Strings in 10d $\mathcal{N}=(1,0)$ supergravity}\label{10d}
Consistent quantum supergravity theories in ten-dimensions are quite limited due to existence of anomalies.
The anomalies of 10d $(1,0)$ supergravity theories can be cancelled by the Green-Schwarz mechanism \cite{Green:1984sg}. The anomaly cancellation allows only 4 choices for gauge groups:
\begin{align}
  SO(32) \ , \quad E_8\times E_8 \ , \quad E_8\times U(1)^{248} \ , \quad U(1)^{496} \ .
\end{align}
See Appendix \ref{appA} for details.

The 10d supergravity theories with the former two gauge groups $SO(32)$ and $E_8\times E_8$ are realized as low energy limits of the type I and the heterotic string theories. On the other hand, it was argued in \cite{Adams:2010zy} that two other theories with abelian gauge factors are not consistent at the quantum level due to anomalies in the context of abelian gauge invariance.


We will now propose a novel stringent condition ruling out the latter two theories with abelian gauge factors by using 2d strings coupled to these 10d theories.
When 2d strings couple to the 10d supergravity, the worldsheet degrees of freedom in general develop local gravitational and gauge anomalies. 
The worldsheet anomalies can be cancelled by the anomaly inflow from the 10d bulk theory toward the 2d strings.
In the following, we will derive the anomaly inflow for 2d strings in the 10d supergravity by employing the method developed in \cite{Callan:1984sa,Blum:1993yd,Freed:1998tg}. We will then check if the anomaly inflow can be cancelled by local anomalies in a unitary worldsheet theory, using the IR properties of the strings and the resulting effective CFT on them. When this cancellation cannot occur, the 10d supergravity becomes an inconsistent theory hosting non-trivial anomalies on the 2d strings.

Strings are sources for the 2-form tensor field $B_2$, which by assumption of completeness of the spectrum in a gravitational theory should exist.  Moreover it is easy to show that they are stable due to the BPS condition.   A string with tensor charge $Q$ adds to the 10d action the tensor coupling
\begin{align}
	S^{\rm str} = Q\int_{\mathcal{M}_{10}} B_2 \wedge \prod_{a=1}^8\delta(x^a)dx^a = Q\int_{\mathcal{M}_2}B \ .
\end{align}
The 2-form $B$ transforms under the local gauge and the local Lorentz symmetry \cite{Bergshoeff:1981um,Chapline:1982ww} (with parameters $\Lambda_i$ and $\Theta$ respectively) as
\begin{align}
	B_2 \ \rightarrow \ B_2 - \frac{1}{4}\sum_i{\rm Tr}(\Lambda_i F_i) + {\rm tr}(\Theta R) \ ,
\end{align}
where $F_i$ denotes the gauge field strengths and $R$ denotes the curvature 2-form of the 10d spacetime. 

The string action $S^{\rm str}$ is not invariant under these local transformations
\begin{align}
	\delta_{\Lambda,\Theta}S^{\rm str} = Q \int_{\mathcal{M}_2}\left[-\frac{1}{4}\sum_i{\rm Tr}(\Lambda_i F_i)+{\rm tr}(\Theta R) \right]\ .
\end{align}
As a consequence the introduction of 2d strings induces an anomaly inflow along the worldsheet of the strings. The anomaly inflow is characterized by the 4-form anomaly polynomial which in this case is given by
\begin{align}
	I_4^{\rm inflow} = Q \left(-\frac{1}{4}\sum_i{\rm Tr} F_i^2 +{\rm tr}R^2\right) \ .
\end{align}
These anomalies must be cancelled by the anomalies coming from the worldsheet degrees of freedom living on the strings.

A half-BPS string coupled to the 10d supergravity gives rise to an $\mathcal{N}=(0,8)$ superconformal field theory (SCFT) at low energy.  To find the chirality of the supersymmetry one uses the condition that we start with a chiral theory in 10d, and for a BPS string we preserve half the supersymmetries, leading to a definite chirality for the supercurrents on the worldsheet.  Supersymmetry on the BPS string also shows that the current for the group has opposite chirality to that of supersymmetry.  We choose conventions so that the supersymmetry current is right-moving and the current for the group is left-moving.

To cancel the anomaly inflow from the bulk gravity theory, the gravitational and the gauge anomalies of the SCFT on a string must be
\begin{align}
	I_4 &= - I_4^{\rm inflow} \nonumber \\
	&= Q\left[\frac{1}{2}p_1(T_2)-c_2(SO(8)) +\frac{1}{4}\sum_i{\rm Tr}F_i^2\right] \ .
\end{align}
Here we used the decomposition
\begin{align}
	{\rm tr}R^2 = -\frac{1}{2}p_1(T_2) +c_2(SO(8)) \ ,
\end{align}
where $p_1(T_2)$ is the first Pontryagin class of the two-manifold $\mathcal{M}_2$ and $c_2(SO(8))$ is the second Chern class of the $SO(8)$ R-symmetry bundle of the worldsheet theory.


Note that the above result involves the contribution from the center of mass degrees of freedom. The center of mass modes form a free $(0,8)$ multiplet $(X_\mu, \lambda_+^I)$ with $\mu,I=1,\cdots,8$ where $X_\mu$ parametrize the motion of strings along 8 transverse directions and $\lambda^I_+$ is the right-moving fermion in the $SO(8)$ spinor representation. From this, we read the anomaly polynomial for the center of mass modes
\begin{align}
	I_4^{\rm com} = -\frac{1}{6}p_1(T_2) - c_2(SO(8)) \ .
\end{align}
So, the anomaly polynomial of the interacting sector in the 2d worldsheet SCFT is given by $I_4' = I_4 - I_4^{\rm com}$.

Let us now focus on the 2d SCFT on a single string, i.e. $Q=1$. The anomaly polynomial of this CFT is 
\begin{align}
	I_4' = I_4-I_4^{com}= \frac{2}{3}p_1(T_2) +\frac{1}{4}\sum_i {\rm Tr}F_i^2 \ .
\end{align}
The left-moving and the right-moving central charges $c_L,c_R$ and the level $k_i$'s of gauge algebras in the worldsheet SCFT can be computed from the anomaly polynomial $I_4'$. The relative central charge $c_R-c_L$ is the coefficient of the gravitational anomaly $-\frac{1}{24}p_1(T_2)$ and the right-moving central charge is $c_R=3k_R$ where $k_R$ is the 't Hooft anomaly coefficient of the superconformal R-symmetry current at the IR fixed point. One finds that 't Hooft anomalies for the $SO(8)$ R-symmetry in $I_4'$ vanish. The level $k_i$ is the coefficient of the gauge anomaly term $\frac{1}{4}{\rm Tr}F_i^2$.
We then compute
\begin{align}
	c_L = 16 \ , \quad c_R = 0 \ , \quad k_i = 1 \ .
\end{align}

The central charges are constrained by unitarity conditions on 2d CFTs, which can be viewed as IR degrees of freedom on the strings.
The central charge realizing the level-$k$ Kac-Moody algebra of group $G$ is (see, e.g., \cite{DiFrancesco:1997nk}):
\begin{align}\label{eq:central-charge-Galgebra}
	c_G = \frac{k \cdot {\rm dim}G}{k+h^\vee} \ ,
\end{align}
where dim$G$ is the dimension and $h^\vee$ is the dual Coxeter number of group $G$ respectively. The central charge for $U(1)$ current algebra is $c_{U(1)}=1$ for any $k_{U(1)}$. For $(0,8)$ SCFTs, the current algebra for group $G$ is on the left-moving sector. This tells us that
\begin{align}\label{eq:10bound}
	\sum_i c_i = \sum_i \frac{k_i\cdot {\rm dim}G_i}{k_i + h_i^\vee} \le c_L \ ,
\end{align}
for a unitary CFT on a string.

We find that the 10d supergravity theories with abelian gauge groups contains non-unitary strings violating this inequality. The $U(1)^{496}$ and $U(1)^{248}$ abelian factors in these theories give rise to too many left-moving modes for the current algebras in the worldsheet CFT, and the central charge of the current algebra exceeds $c_L=16$, namely $\sum_ic_i>c_L$. Therefore we conclude that 10d supergravity theories with $U(1)^{496}$ and
$E_8\times U(1)^{248}$ gauge groups are inconsistent when coupled to 2d strings, and thus they belong to the swampland. On the other hand, the central charges on a single string in the 10d supergravities with $SO(32)$ or $E_8\times E_8$ gauge group
saturate the bound (\ref{eq:10bound}) as $\sum_ic_i=c_L=16$, so the string can consistently couple to these 10d theories.

\section{Strings in 6d $\mathcal{N}=(1,0)$ supergravity}\label{6d}
We now turn to six-dimensional supergravity theories preserving 8 supersymmetries.  There are four kinds of massless supermultiplets appearing in such theories: a gravity multiplet, tensor multiplets, vector multiplets, and hypermultiplets. 6d supergravity theories may have anomalies, which are characterized by an 8-form anomaly polynomial $I_8$, from the chiral fields in these multiplets.

Let us consider a gravity theory coupled to $T$ tensor multiplets and vector multiplets of the gauge group $G=\prod_iG_i$, as well as hypermultiplets transforming in representation ${\bf R}$ of the gauge group. The chiral fields such as the self-dual and anti-self dual two-forms $B_{\mu\nu}^\pm$, a gravitino,  and other chiral fermions in this theory contribute to the anomalies for the gauge and Lorentz transformations. Such anomalies can exactly be computed by evaluating 1-loop box diagrams for the chiral fields with four external gravitational and gauge sources. 
Consistent quantum supergravity theories must be free of such anomalies. Thus non-vanishing 1-loop anomalies must be cancelled for the 6d theories that are consistent at the quantum level, which leads to quite stringent constraints.

The 1-loop anomalies can be cancelled by the Green-Schwarz-Sagnotti mechanism \cite{Sagnotti:1992qw} if the anomaly polynomial factorizes as
\begin{align}\label{eq:X4}
  I_8^{1-loop} &= \frac{1}{2}\Omega_{\alpha\beta}X_4^\alpha X_4^\beta \ , \nonumber \\
  X_4^\alpha &= \frac{1}{2}a^\alpha {\rm tr} R^2 + \frac{1}{4}\sum_i b_i^\alpha \frac{2}{\lambda_i}{\rm tr} F_i^2 \ ,
\end{align}
where $\Omega_{\alpha\beta}$ is a symmetric bilinear form of $T+1$ tensors with a signature $(1,T)$, and $a^\alpha$ and $b_i^\alpha$ are vectors in $\mathbb{R}^{1,T}$. 

\begin{table}[t]
\centering
\begin{tabular}{|c|c|c|c|c|c|c|c|c|}
	\hline
	$G$ & $SU(N)$ & $SO(N)$ & $Sp(N)$ & $G_2$ & $F_4$ & $E_6$ & $E_7$ & $E_8$ \\
	\hline 
	$\lambda$ & $1$ & $2$ & $1$ & 2 & 6 & 6 & 12 & 60 \\ 
	\hline
\end{tabular}
\caption{Group theory factors}\label{tb:G-factors}
\end{table}

The conditions for the factorization can be summarized as follows:
\begin{align}\label{eq:6dconstraints}
  &H-V = 273 - 29T \ , \quad a\cdot a = 9-T \ , \nonumber \\
  &0= B^i_{\bf adj} - \sum_{\bf R} n^i_{\bf R}B^i_{\bf R} \ , \nonumber \\
  &a\cdot b_i = \frac{\lambda_i}{6}\left(A^i_{\bf adj} - \sum_{\bf R} n^i_{\bf R} A^i_{\bf R}\right) \ , \nonumber \\
  &b_i\cdot b_i = \frac{\lambda_i^2}{3}\left(\sum_{\bf R} n^i_{\bf R} C^i_R - C^i_{\bf adj}\right)  \ , \nonumber \\
  &b_i\cdot b_j = 2\lambda_i\lambda_j\sum_{\bf R,S}n^{ij}_{\bf R,S}A^i_{\bf R} A^j_{\bf S} \quad (i\neq j) \ ,
\end{align}
where $\Omega_{\alpha\beta}$ is used for the inner product of two vectors, like $v\cdot w=\Omega_{\alpha\beta} v^\alpha w^\beta$.
Here $V$ and $H$ are the number of vector and hyper multiplets, and $n_{\bf R}^i$ denotes the number of hypermultiplets in the representation ${\bf R}$ for gauge group $G_i$ and $A^i_{\bf R},B^i_{\bf R},C^i_{\bf R}$ are group-theory factors for each representation defined as follows:
\begin{align}
  {\rm tr}_{\bf R}F^2 = A_{\bf R}{\rm tr}F^2 \ , \quad {\rm tr}_{\bf R}F^4 = B_{\bf R}{\rm tr}F^4 + C_{\bf R}({\rm tr}F^2)^2 \ .
\end{align}
When these conditions are satisfied, the perturbative anomaly factorizes and it can be cancelled by adding to the action the Green-Schwarz term
\begin{align}
  S_{GS} = \int  \Omega_{\alpha\beta}B_2^{\alpha}\wedge X_4^\beta \ .
\end{align}
This term induces tree-level anomalies of the form $I_8^{GS}=-\frac{1}{2}\Omega_{\alpha\beta}X_4^\alpha X_4^\beta$ that exactly cancels the factorized anomaly $I_8^{1-loop}$.
So, 6d supergravity theories satisfying the conditions in equation (\ref{eq:6dconstraints}) have no apparent quantum anomalies and seem to be consistent. Extensive lists of would-be consistent 6d supergravity theories are given in various literature \cite{Schwarz:1995zw, Kumar:2009us, Kumar:2009ae, Kumar:2010ru, Kumar:2010am, Johnson:2016qar, Taylor:2018khc, Raghuram:2018hjn, Taylor:2019ots}
(see \cite{Taylor:2011wt} for a review).

\subsection{Central charges of 2d $(0,4)$ SCFTs on strings}
Let us now consider 2d strings in 6d supergravity theory without manifest anomalies. We will discuss additional conditions from the 6d/2d coupled system. Strings are sources for the two-form fields $B^\alpha_2$ and thus should exist by assumption of completeness of the spectrum in a gravitational theory. We shall consider BPS strings preserving half supersymmetries. The worldsheet theory on those strings gives rise to 2d $(0,4)$ SCFT at low energy.
 As discussed in the 10d cases, the degrees of freedom living on the string worldsheet can have non-zero anomalies and these anomalies must be cancelled through the anomaly inflow mechanism. The anomaly inflow in 6d SCFTs was studied in \cite{Kim:2016foj,Shimizu:2016lbw} (See also \cite{Hayashi:2019fsa} for generalization to 6d supergravities from F-theory compactification). See Appendix \ref{appB} for a brief review on the anomaly inflow to 2d strings in 6d SCFTs and 6d supergravity theories. 
%

The 2d SCFT on strings with charge $Q^\alpha$ in the 6d supergravity theory has the anomaly polynomial of this form 
\begin{align}
  I_4 &= \Omega_{\alpha\beta}Q^\alpha\left(\frac{1}{2}a^\alpha {\rm tr}R^2 + \frac{1}{4}\sum_i b^\alpha_i {\rm Tr}F_i^2 +\frac{1}{2} Q^\beta\chi_4(N_4)\right) \nonumber \\
  & = -\frac{Q\cdot a}{4}p_1(T_2)+ \frac{1}{4}\sum_i Q\cdot b_i {\rm Tr}F_i^2 \nonumber \\
  &\qquad  -\frac{Q\cdot Q-Q\cdot a}{2}c_2(R) + \frac{Q\cdot Q+Q\cdot a}{2}c_2(l) \ .
\end{align}
In this computation, we used the decomposition ${\rm tr}R^2 = -\frac{1}{2}p_1(T_2) +c_2(l) +c_2(R)$.

This result involves the contribution from the center of mass degrees of freedom which decouples in the IR SCFT. The center of mass modes consist of 4 bosons common to left- and right-movers and 4 right-moving fermions and they form a free hypermultiplet $(X_{a\dot{a}},\lambda_a)$ where $a,\dot{a}$ are indices for $SU(2)_l\times SU(2)_R$. They contribute to the anomaly as
\begin{align}
  I_4^{com} = -\frac{1}{12}p_1(T_2) -c_2(l) \ .
\end{align}
Therefore the anomaly polynomial of the 2d worldsheet theory after removing the center of mass contributions becomes
\begin{align}
  I_4'&= I_4 - I_4^{com} \nonumber \\
  &=-\frac{3Q\cdot a\!-\!1}{12}p_1(T_2)+ \frac{1}{4}\sum_i Q\cdot b_i {\rm Tr}F_i^2 \nonumber \\
  & \quad -\frac{Q\cdot Q\!-\!Q\cdot a}{2}c_2(R) + \frac{Q\cdot Q\!+\!Q\cdot a\!+\!2}{2}c_2(l) \ .
\end{align}

The central charges of the 2d SCFT can be extracted from the anomaly polynomial as discussed in the previous section. The relative central charges $c_R-c_L$ is again the coefficient of the gravitational anomaly. The right-moving central charge $c_R$ is associated to the anomaly coefficient of the R-symmetry current. Here, we should be careful about the R-symmetry at the IR fixed point. It is possible that an accidental symmetry emerges at low energy and it takes over the role of the R-symmetry in the IR $(0,4)$ superconformal algebra. It is also possible that a 2d worldsheet theory degenerates to a product of distinct SCFTs carrying different IR R-symmetries.

Indeed, this happens for the strings in local 6d SCFTs or little string theories (LSTs) embedded in the supergravity theories. The 2d SCFTs on such strings have an accidental $SU(2)_I$ symmetry in the decouping limit and this symmetry becomes the $SU(2)$ R-symmetry in the $(0,4)$ superconformal algebra. This $SU(2)_I$ is descended from the $SU(2)$ R-symmetry of the local 6d SCFTs or LSTs, but it is broken in the full supergravity theory. The free theory with the center of mass degrees of freedom we discussed above also has the same accidental $SU(2)_I$ symmetry.

It is therefore crucial to identify the right R-symmetry in the IR SCFTs. Only after this we can extract the correct central charges in the IR SCFTs. From now on we will focus on the strings in the 6d supergravity theory that give rise to a single interacting SCFT at low energy without the accidental $SU(2)_I$ symmetry. The IR SCFTs on such supergravity strings (not strings in local 6d SCFTs or LSTs) have the $(0,4)$ superconformal algebra with an $SU(2)_R$ R-symmetry. The conditions for this type of strings will be given below.
The right-moving central charge $c_R$ of these SCFTs can then be read off from the anomaly coefficient of the $SU(2)_R$ symmetry.
For a non-degenerate 2d SCFT on a supergravity string, the central charges $c_L,c_R$ are given by
\begin{align}\label{eq:central-2d}
  c_L = 3Q\cdot Q-9Q\cdot a +2 \ , \quad c_R = 3Q\cdot Q - 3Q\cdot a\ .
\end{align}
The central charges $k_i$ and $k_l$ for the bulk gauge symmetries $G_i$ and $SU(2)_l$ can also be extracted from the anomaly polynomial. We find
\begin{align}\label{eq:central-2d-k}
  k_i = Q\cdot b_i \ , \quad k_l = \frac{1}{2}(Q\cdot Q+Q\cdot a+2) \ .
\end{align}

A large class of 6d $(1,0)$ supergravity theories can be engineered in F-theory on elliptic Calabi-Yau 3-folds. In the context of F-theory, the 2d SCFT with string charge $Q$ arises as a low energy theory on a D3-brane wrapping genus $g$ curve $C=Q$ in the base $B$ of the 3-fold.
We can compare the above results against the central charges of the strings coming from D3-branes in F-theory. The 2d SCFT for a D3-brane wrapping a genus $g$ curve $C$ inside $B$ has the central charges \cite{Haghighat:2015ega} (See also \cite{Hayashi:2019fsa})
\begin{align}
	c_L' = 3C\!\cdot\! C-9K\!\cdot\! C+6 \ , \quad c_R' = 3C\!\cdot\! C -3 K\!\cdot\! C +6 \ ,
\end{align}
where $K$ is the canonical class of $B$, and it has a $SU(2)_l$ current algebra at level $k_l' = g-1$.
Here the genus $g$ of the curve $C$ can be computed by the Riemann-Roch theorem
\begin{align}
	C\cdot C +K\cdot C = 2g-2 \ .
\end{align}

These results again include the contribution from the center of mass modes; 4 left- and 4 right-moving bosons and 4 right-moving fermions. The central charges of the center of mass modes are $c_L^{com}=4,c_R^{com}=6$ and, as discussed in \cite{Haghighat:2015ega}, they contribute to the $SU(2)_l$ current algebra by $k_l^{com}=-1$.

One can easily see that the central charges $c_L',c_R',k_l'$ in F-theory models after removing the center of mass contributions are in perfect agreement with the central charges of 2d SCFTs from the anomaly inflow given in (\ref{eq:central-2d}) and (\ref{eq:central-2d-k}). To see this agreement, one needs to identify the inner product $\Omega$ among tensors with the intersection form in $H_2(B,\mathbb{Z})$, and map the vector $a$ to the canonical class $K$ in the base of the elliptic CY$_3$. 
This comparison confirms our anomaly inflow computation for 2d strings in 6d supergravity theories.

\subsection{Consistency conditions}

We shall now show that the consistency of 2d worldsheet theories encoded in the central charges imposes additional conditions on 6d supergravity theories.

Let us consider the moduli space of a 6d supergravity theory that is parametrized by scalar fields in the tensor multiplets as well as the scalar field in the hypermultiplet controlling the overall volume of the tensor moduli space.  From supergravity considerations, for this moduli space being well-defined,  we should be able to find a linear combination of these scalar fields, which we call $J$, satisfying
\begin{align}
	J\cdot J> 0 \ , \quad J\cdot b_i >0 \ , \quad -J\cdot a >0 \ .
\end{align}
This $J$ plays the role of the central charge in the supersymmetry algebra for the $B$-fields.
The first condition stands for the metric positivity of the tensor branch along $J$. The second one is the condition for the gauge kinetic term along $J$ to have proper sign on the tensor moduli \cite{Sagnotti:1992qw}. Otherwise, the gauge kinetic term has a wrong sign and it leads to an instability. 
The last condition ensures, through supersymmetry, the positivity of the Gauss-Bonnet term in gravity \footnote{The coefficients of the Gauss-Bonnet term, the Riemann-squared term, and the Weyl-squared term are all equal. This can be seen using the equations of motion to rewrite the operators involving $R$ and $R_{\mu \nu}$.}.
While there have been attempts to prove the positivity of the curvature-squared corrections in $D>4$ using e.g. unitarity \cite{Cheung:2016wjt}, 
the singular UV behavior due to graviton exchange prevents one from making such spectral decomposition argument \cite{Hamada:2018dde}. Here, we note that even if we impose this last condition, there seem to be infinitely many anomaly-free 6d supergravity theories (see \cite{Taylor:2011wt} for a review). We thus assume its validity, leaving a derivation for future work.

In F-theory realization \footnote{This last condition in F-theory setup translates to the condition that $J\cdot K <0$, signifying that the base of F-theory compactification is positively curved, which is necessary for solving Einstein's equation when $\tau$ varies over the base.}, this combination $J$ corresponds to a K\"ahler form $J\in H^{1,1}(B)$ of the base $B$. The above conditions on $J$ define a positive-definite K\"ahler cone on $B$. We will call $J$ a K\"ahler form for all 6d theories regardless of whether it has an F-theory realization.

The tensions of 2d BPS strings are determined with respect to the K\"ahler form $J$. This imposes a condition $Q\cdot J \ge 0$ on the string charge $Q$. A worldsheet theory has non-negative tension only if $Q\cdot J \ge 0$.

The strings with $Q\cdot J \ge 0$ embedded in 6d supergravity theories must give rise to unitary 2d SCFTs. For a unitary 2d CFT, the central charges must be non-negative, i.e. $c_L,c_R\ge0$. If the central charges computed through the anomaly inflow for a string are negative, the corresponding anomalies cannot be cancelled by a unitary 2d worldsheet theory. This results in that the 6d supergravity theory with such strings is inconsistent hosting non-vanishing anomalies along the 2d string worldsheet, and it thus belongs to the swampland. So we can use the anomaly inflow on 2d strings to analyze the consistency of 6d supergravity theories.

We remark that the strings in 6d SCFTs or little string theories (LSTs) contained in 6d supergravity theories in general lead to 2d CFTs having a negative value for $c_R$ given in (\ref{eq:central-2d}).
For example, the unit string charge $Q$ for a 2d string in the 6d $SO(8)$ non-Higgsable SCFT have the properties $Q\cdot Q=-4$ and $Q\cdot K=+2$. So the value for $c_R$ of this string with unit charge $Q$ is $-18$.
 This seems to say that the theory is inconsistent since its central charge is negative $c_R<0$ by the formula in (\ref{eq:central-2d}).
However, this is not the case. Note that the central charge $c_R$ above is obtained by assuming the R-symmetry of the low energy $(0,4)$ SCFT is the $SU(2)_R$. As discussed, the strings in local 6d SCFTs or LSTs have an accidental $SU(2)_I$ symmetry and this becomes the R-symmetry of the low energy SCFT. Therefore $c_R$ in such strings is different from what we computed above. The central charges of various worldsheet theories in 6d SCFTs are computed in the literature \cite{Kim:2016foj,Shimizu:2016lbw}, and one can check that those theories have positive central charges $c_R,c_L$ with respect to the $SU(2)_I$ R-symmetry.


We are interested in the configurations of a single string in the 6d supergravity that have $SU(2)_R$ as the R-symmetry in the superconformal algebra and that do not degenerate to a product of disconnected 2d SCFTs at low energy. A single string state has no bosonic zero mode along the transverse $\mathbb{R}^4$ directions except the center of mass degrees of freedom. This implies that, after removing the center of mass modes, the worldsheet theory on a string contains the $SU(2)_l$ current algebra realized on the left-movers. So the $SU(2)_l$ central charges should be non-negative, i.e. $k_l\ge0$. In F-theory compactification, this condition becomes a trivial condition saying that $g\ge0$  for a string wrapped on a genus $g$ curve $Q$.
The central charge conditions $c_R\ge0$ and $k_l\ge0$  on these SCFTs can be summarized as
\begin{align}\label{eq:central-charge-conditions}
  Q\cdot Q \ge -1 \ , \quad Q\cdot Q + Q\cdot a \ge -2 \ .
\end{align}

There are more conditions associated to the flavor central charges $k_i=Q\cdot b_i$. The flavor central charge measures the index of the bulk fields charged under the gauge group $G_i$ on the string background with charge $Q$. So it counts the number of zero modes at the intersection between the tensor carrying the gauge group $G_i$ and the tensor labelled by the string charge $Q$. Unless the string degenerates to an instanton string of the group $G_i$, namely unless $Q\sim b_i$, the flavor central charge can receive  contributions only from fermionic zero modes which are in the left-moving sector.  This means that the flavor central charges of the 2d SCFTs on non-degenerate strings (not in local 6d SCFTs or LSTs) in 6d supergravities should be non-negative. In other words,
\begin{align}\label{eq:condition-G}
  k_i = Q\cdot b_i \ge 0 \ ,
\end{align}
for the strings we are interested in, where we used the convention that left-movers have positive contributions to flavor central charges. In the F-theory viewpoint, the condition (\ref{eq:condition-G}) is the same as the condition that the curve class $Q$ is effective and irreducible within the Mori cone of the K\"ahler base $B$.

Note that a 2d theory on instanton strings can have right-movers associated to bosonic zero modes parametrizing the moduli space of $G_i$ instantons. These right-movers can provide negative contributions to the flavor central charges.
However, such instanton strings correspond to the strings in local 6d SCFTs or 6d LSTs. When a string degenerate to a product of the instanton strings, the low energy theory will include 2d theories for the strings in local 6d SCFTs or LSTs which have the accidental $SU(2)_I$ R-symmetry. As discussed above, we are not interested in the worldsheet theories with $SU(2)_I$ R-symmetry.
So we shall only focus on strings and the associtated 2d SCFTs satisfying the condition (\ref{eq:condition-G}) as well as (\ref{eq:central-charge-conditions}).

For such 2d SCFTs, we have $G_i$ current algebra with level $k_i$. 
Using supersymmetry algebra in the context of BPS strings, one can show that the current algebra is on the left-movers in the $(0,4)$ SCFTs and its central charge contribution is given in (\ref{eq:central-charge-Galgebra}). Therefore, we find the following constraint on the 2d worldsheet SCFT in the 6d supergravity:
\begin{align}\label{eq:c-inequality}
	\sum_i \frac{k_i\cdot {\rm dim}G_i}{k_i+h_i^\vee} \le c_L \ .
\end{align}
So the 2d SCFTs on strings satisfying the conditions in the equations (\ref{eq:central-charge-conditions}) and (\ref{eq:condition-G})
must have central charges constrained by the equation (\ref{eq:c-inequality}). Otherwise, the 2d worldsheet theory is non-unitary. In conclusion, we claim that a 6d supergravity theory embedding 2d strings whose worldsheet theory violates the condition (\ref{eq:c-inequality}) is inconsistent and it therefore belongs to the swampland.

\subsection{Examples}
The basic structure of our examples is as follows.  For each one we have the $\Omega, a,b_i$ given by anomaly cancellation conditions.
We use this to find the allowed ranges for $J$ and choose a particular $J$ in the allowed region.  We then use this to restrict the allowed string charges $Q$'s and use that to compute central charges $c_R,c_L$ and $k_l,k_i$ and see if we have any contradictions with unitarity.

Let us first consider the 6d supergravity theory coupled to $T=9$ tensors with $SU(N)\times SU(N)$ gauge group and two bi-fundamental hypermultiplets introduced in \cite{Kumar:2010ru} (See also \cite{Schwarz:1995zw} for $T=1$ models). The anomaly polynomial of this model factorizes for an arbitrary $N$ and hence it seems that they provide an infinite family of consistent 6d supergravity theories. It was however shown in \cite{Kumar:2010ru} that these models have no F-theory realization at large enough $N$.

Let us examine these models with 2d strings to see if the consistency conditions of the worldsheet theory on the strings can provide any bound on $N$.

We can always choose a tensor basis such that the bilinear form $\Omega$ and the vectors $a,b_1,b_2$ are given as follows \cite{Kumar:2010ru}:
\begin{align}
  \Omega &= {\rm diag}(+1,(-1)^9) \ , \quad a = (-3,(+1)^9) \ , \nonumber \\
  b_1 &= (1,-1,-1,-1,0^6) \ , \quad b_2 = (2,0,0,0,(-1)^6) \ .
\end{align}
In this basis, one can easily see that a K\"ahler form chosen as $J=(1,0^9)$ satisfies the conditions $J^2>0, J\cdot b>0$ and $J\cdot a<0$.

Consider a string of a generic charge $Q=(q_0,q_1,\cdots,q_9)$ with $q_i\in \mathbb{Z}$. This string with $q_0>0$ has a positive tension with respect to $J$. The conditions (\ref{eq:central-charge-conditions}) and (\ref{eq:condition-G}) on the IR SCFT for this string can be summarized as
\begin{align}\label{eq:NxN-strings}
    &q_0^2 -\sum_{i=1}^9q_i^2 \ge -1 \ ,\quad  q_0^2 -\sum_{i=1}^9q_i^2 -3q_0-q_{1:3}-q_{4:9} \ge -2 \ , \nonumber \\
    &k_1 = q_0+ q_{1:3}\ge 0 \ , \quad k_2 = 2q_0 + q_{4:9}\ge0 \ ,
\end{align}
where $q_{1:3}\equiv \sum_{i=1}^3q_i$ and $q_{4:9}\equiv\sum_{i=4}^9q_i$.
In addition, the flavor central charges are restricted by the unitarity bound (\ref{eq:c-inequality})
\begin{align}\label{eq:bound-NxN}
  \frac{k_1(N^2-1)}{k_1+N}+\frac{k_2(N^2-1)}{k_2+N} \le c_L \ ,
\end{align}
where the left-moving central charges is
\begin{align}
  c_L = 3(q_0^2 - \sum_{i=1}^9q_i^2) + 9(3q_0+q_{1:3}+q_{4:9}) + 2 \ .
\end{align}
As discussed above, if this bound is violated for any $Q$ satisfying (\ref{eq:NxN-strings}), the anomaly inflow from the bulk 6d supergravity theory cannot be cancelled by a unitary 2d CFT which renders the 6d supergravity inconsistent at the quantum level.

The bound (\ref{eq:bound-NxN}) gives the strongest constraint on $N$ of the 6d supergravity theory  when the left-hand side is maximized, namely $k_i$'s are minimized, while the right-hand side is minimized. This implies the strongest bound can be given by a string with $q_0^2-\sum_iq_i^2=-1$ and $k_1=0,k_2=1$. This occurs for $Q=(1,-1,0,0,-1,0^5)$. The central charge bound for the string configuration being unitary is
\begin{align}
 \frac{k_2(N^2-1)}{k_2+N}\le c_L \ \rightarrow \ \frac{N^2-1}{1+N} \le 8 \ \rightarrow \ N\le9\ .
\end{align}
Therefore the 6d supergravity theory with $N>9$ belongs to the swampland containing non-unitary string configurations. This bound is stronger than the bound  $N\le12$ from the Kodaira condition in F-theory \cite{Kumar:2010ru}.  It is interesting that we can thus rule out would be purely geometric constructions which could have in principle realized this model for $N=10,11,12$.  In other words our arguments can be used to teach us some facts about the geometry of elliptic Calabi-Yau threefolds!
Also, it is reassuring that this bound does not rule out the string theory realization for $N=8$ given in \cite{Dabholkar:1996zi,Dabholkar:1996pc} and all the $N\leq 8$ theories which one can obtain from it by partial Higgsing. Remarkably, our worldsheet analysis provides a new bound on the rank of gauge groups in the 6d bulk supergravity theory and the result is consistent with the F-theory argument and also the known string theory realization.  It would be interesting to see if one can construct the $N=9$ case which we were not able to rule out.

The second example is the 6d supergravity with $T=1$ and $SU(N)$ gauge group coupled to one symmetric and $N-8$ fundamental hypermultiplets first introduced in \cite{Kumar:2009ac,Kumar:2010ru}. 
The rank of the gauge group is bounded as $N\le 30$ from the 6d anomaly cancellation conditions. For this model, we are free to choose a tensor basis giving
\begin{align}
  \Omega = {\rm diag}(1,-1) \ ,\quad a = (-3,1) \ , \quad b=(0,-1) \ .
\end{align}
The K\"ahler form can always be chosen as $J=(n,1)$ with $n^2>1$ and $n>0$.
This theory has no F-theory realization because, when we identify the base $B$ with a Hirzebruch surface $\mathbb{F}_1$, the tensor for $b$ cannot be mapped to any effective curve class \cite{Kumar:2010ru}.

We shall now see if the consistency conditions on string configurations of this 6d theory can provide a stronger bound on the rank $N$. Consider a generic string with $Q=(q_1,q_2)$ satisfying the conditions (\ref{eq:central-charge-conditions}), (\ref{eq:condition-G}), namely
\begin{align}
  &q_1^2-q_2^2 \ge -1 \ , \quad q_1^2-q_2^2-3q_1-q_2\ge -2 \ , \nonumber \\
  &k = Q\cdot b = q_2 \ge 0 \ .
\end{align}
Also, $nq_1>q_2$ from $J\cdot Q>0$. These conditions can be then simplified, for the strings interacting with the gauge group, as
\begin{align}
  q_1 \ge 3 \ \quad q_1-2 \ge q_2 >0 \ .
\end{align}
The constraint on the central charges
\begin{align}
  \frac{q_2(N^2-1)}{q_2+N} \le 3(q_1^2-q_2^2)+9(3q_1+q_2)+2 
\end{align}
can provide the strongest bound on $N$ when $Q=(3,1)$, and the bound is $N\le 117$. This bound is weaker than the bound $N\le30$ coming from the 6d anomaly cancellation conditions. This may imply, unless another inconsistency is revealed by any other means, that these 6d supergravity models with $N\le 30$ are all consistent theories though they do not seem to admit an F-theory realization.

The anomaly inflow consideration can provide a new bound on a family of models with $T=8k+9$ and gauge group $G=(E_8)^k$ for arbitrary large $k$, which was introduced in \cite{Kumar:2010ru}. The vectors $a$ and $b_i$ in the anomaly polynomial satisfy $a\cdot b_i=10,\ b_i\cdot b_j=-2\delta_{ij}$ with $i,j=1,\cdots,k$.
When $k\ge3$, one can choose a basis for tensors in \cite{Kumar:2010ru} that gives rise to
\begin{align}\label{eq:basis-E8}
	\Omega &= {\rm diag}(1,(-1)^{8k+9}) \ , \quad a = (-3,1^{8k+9}) \ , \nonumber\\
	b_i &= (-1,-1,0^{4(i-1)},(-1)^3,-3,0^{8k+8-4i}) \ , 
\end{align}
The K\"ahler form in this basis can be chosen as
\begin{align}
	J = (-j_0,0^{4k+1},1^{4k+8})\ , \ \ (4k+8)/3> j_0 >\sqrt{4k+8} \ .
\end{align}

Now consider a string with charge $Q=(-q,0^{8k+9})$ in this 6d model. This string has a positive tension if $q>0$. Moreover, the conditions $k_l\ge0,c_R\ge0$ and $k_i\ge0$ can be satisfied if $q>2$. However, the bound on the levels of flavor current algebras $k_i =Q \cdot b_i = q$:
\begin{align}
\sum_{i=1}^k \frac{248 k_i}{k_i+30} \le c_L \ \ \  \rightarrow \ \ \  k \frac{248q}{q+30} \le 3q(q-9)+2
\end{align}
cannot be satisfied by, for example, strings with charge $3\le q\le14$ for any $k\geq 3$. This result demonstrates that all these 6d supergravity models for $k\ge3$ endowed with the bilinear form $\Omega$ and vectors $a,b_i$ given in (\ref{eq:basis-E8}) reveal non-vanishing anomalies on the 2d strings, and therefore they are in the swampland.

Note however that the 6d supergravity theories of this type for $k\le2$ are not ruled out by this  analysis. When $k=1,2$, there exists another solutions of $\Omega$ and $a,b_i$ cancelling the anomalies, like this:
\begin{align}
	\Omega &= {\rm diag}(1,(-1)^{17}) \ , \quad a = (-3,1^{17}) \ , \nonumber\\
	b_1 &= (0,1,(-1)^{11},0^{5}) \ ,
\end{align}
for $k=1$ and
\begin{align}
	\Omega &= {\rm diag}(1,(-1)^{25}) \ , \quad a = (-3,1^{25}) \ , \nonumber\\
	b_1 &= (0,1,(-1)^{11},0^{13}) \ , \quad b_2 = (0,0^{13},1,(-1)^{11}) \ ,
\end{align}
for $k=2$.
Thus the above analysis does not apply to the $k=1,2$ cases.
We 
do not
find any string configuration showing inconsistencies for these cases. Indeed, the 6d gravity theory with $k=2$ can be realized by the compactification of M-theory on $K3\times (S^1/\mathbb{Z}_2)$,  where we place 24 M5 branes on the interval \cite{Seiberg:1996vs}.

The last example is the 6d supergravity theory with $T=0$ and gauge group $SU(8)$ coupled to an exotic  hypermultiplet in the `box' representation, which was introduced in \cite{Kumar:2010am}. This theory cannot be realized in F-theory. The 6d anomaly cancellation sets the vectors as $a=-3$ and $b=8$.

The 2d SCFTs on a string with charge $Q>0$ in this theory satisfy the conditions (\ref{eq:central-charge-conditions}) and (\ref{eq:condition-G}). The strongest constraint on the left-moving central charge is given by the minimal string with $Q=1$. The central charge constraint for this model is marginally satisfied as
\begin{align}
	\frac{k\times 63}{k+8}\le c_L \ \ \rightarrow \ \ 31.5 \le 32 \ \ \ {\rm for} \ k=Q\cdot b=8 \ .
\end{align}
Therefore at least as far as the unitarity constraint is concerned this theory is not ruled out and the strings can consistently couple to this 6d supergravity theory.

\section{Conclusions}\label{conclusions}

In summary, we have discussed the consistencies of 10d and 6d $\mathcal{N}=(1,0)$ supergravity theories 
as seen from 2d strings that couple to the 2-forms in the bulk.
We have identified the central charges of the worldsheet SCFTs on the strings using the anomaly inflow from the bulk supergravity theory.
The unitarity of the worldsheet SCFTs associated to the central charges leads to novel constraints on the allowed supergravity models, that are not visible from the particle viewpoint.

In this paper, we analyzed only a handful of 6d supergravity models.
A large class of would-be consistent 6d supergravity theories has been discussed in the literature, for example \cite{Kumar:2009ac,Kumar:2010ru,Kumar:2010am}. It might be possible to similarly rule out many such models using more detailed constraints from string probes that we considered in this paper. We leave this for future work.

It would be straightforward to generalize the anomaly inflow consideration discussed in this paper to other type of branes coupled to the supergravity theories. Our discussion in this paper is merely a starting point of a bigger program to understand the consistency of quantum gravitational theories in various dimensions by coupling them to all possible branes and defects of the theories. We hope this program ultimately provides a complete classification of consistent supergravity theories in six- and perhaps also other dimensions, and more broadly deepens our understanding of the swampland criteria.

\section{Acknowledgments} 

We would like to thank E.~Bergshoeff, M.~Ro\v cek and W.~Taylor for useful and informative discussions. H.K. and G.S. would like to thank Harvard University for hospitality during part of this work.
H.K. is supported by the POSCO Science Fellowship of POSCO TJ Park Foundation and the National Research Foundation of Korea (NRF) Grant 2018R1D1A1B07042934. G.S. is supported in part by the DOE grant DE-SC0017647 and the Kellett Award of the University of Wisconsin.  The research of CV is supported in part by the NSF grant
PHY-1719924 and by a grant from the Simons Foundation (602883, CV).

\appendix

\section{\\ Anomalies in 10d supergravity theories}\label{appA}

We adopt the normalization used in \cite{Bilal:2008qx}, but a factor of $1/4\pi$ is included in the curvature 2-form $R$ and the field strength $F$ includes a factor of $1/2\pi$. 
An $\mathcal{N}=1$ supergravity theory in ten dimensions contains a Majorana-Weyl gravitino, some spin $\frac{1}{2}$ fermions with negative chirality, and gauginos with positive chirality. The gravitino contributes to the anomaly as
\begin{align}
  I_{12}^{3/2} = -\frac{11}{126} {\rm tr} R^6 +\frac{5}{96}{\rm tr} R^4\, {\rm tr} R^2-\frac{7}{1152}({\rm tr}R^2)^3 \ ,
\end{align} 
while the contribution from a spin $\frac{1}{2}$ fermion is
\begin{align}
  I_{12}^{1/2} &= (tr_{\mathcal{R}}1)
  \left[\frac{1}{5670}{\rm tr}R^6 + \frac{1}{4320}{\rm tr}R^4\, {\rm tr}R^2 \right. \nonumber \\
  & \left. +\frac{1}{10368}({\rm tr}R^2)^3\right]  -\frac{1}{2}tr_{\mathcal{R}}F^2\left[\frac{1}{360}{\rm tr}R^4+\frac{1}{288}({\rm tr}R^2)^2\right] \nonumber \\
  &+\frac{1}{288}(tr_{\mathcal{R}}F^4){\rm tr} R^2-\frac{1}{720}tr_{\mathcal{R}}F^6 \ ,
\end{align}
where $\mathcal{R}$ denotes the representation of the fermion under the gauge algebra.
The total 1-loop anomaly of the theory is given by the sum over all fermion contributions as
\begin{align}
  I_{12}^{1-loop} &= I_{12}^{3/2} - I_{12}^{1/2}|_{\mathcal{R}=1} + I_{12}^{1/2}|_{\mathcal{R}=adj} \nonumber \\
  &= \frac{{\rm dim}G-496}{5670}{\rm tr} R^6 + \frac{{\rm dim}G+224}{4320}{\rm tr} R^4\, {\rm tr} R^2 \nonumber \\
  &+ \frac{{\rm dim}G-64}{10368}({\rm tr} R^2)^3  \nonumber \\
  & -\frac{1}{2}tr_{\rm adj}F^2\left[\frac{1}{360}{\rm tr} R^4+\frac{1}{288}({\rm tr} R^2)^2\right]  \nonumber \\
  &+\frac{1}{288}tr_{\rm adj}F^4\, {\rm tr} R^2-\frac{1}{720}tr_{\rm adj}F^6 \ ,
\end{align}
where  dim$G$ is the dimension of the gauge group. When this 1-loop anomaly factorizes as
\begin{align}
	I_{12}^{1-loop}= X_4\wedge X_8 \ ,
\end{align}
it can be cancelled by the Green-Schwarz mechanism \cite{Green:1984sg}. This factorization condition allows only 4 choices of gauge groups: $SO(32)$, $E_8\times E_8$, $E_8\times U(1)^{248}$, $U(1)^{496}$.

To cancel the 1-loop anomaly, we add to the action the Green-Schwarz term
\begin{align}
	S^{GS} = \int B_2 \wedge X_8 \ .
\end{align}
Here the 2-form field $B_2$ in the 10d theory transforms under local gauge and Lorentz group as
\begin{align}
	B_2\ \rightarrow \ B_2- \frac{1}{4}{\rm Tr}(\Lambda F ) + {\rm tr}(\Theta R) \ ,
\end{align}
where $\Lambda$ and $\Theta$ are the transformation parameters. It then follows that the Green-Schwarz term induces anomalies under the gauge and Lorentz transformations which may cancel the 1-loop anomalies.
We normalize `Tr' such that the integral of $\frac{1}{4}{\rm Tr}F^2$ over a 4-manifold gives the instanton number $Q\in \mathbb{Z}$. Note that the gauge transformation of $B_2$ is fixed by supersymmetry and the gauge invariance of the 3-form field strength $H_3$ \cite{Bergshoeff:1981um,Chapline:1982ww}. The Lorentz transformation of $B_2$ is on the other hand fixed by the higher order correction on $H_3$ in the derivative expansion.

\section{Anomaly inflows from 6d to 2d}\label{appB}

Let us briefly review the anomaly inflow computation in 6d theories in the presence of 2d strings discussed in \cite{Kim:2016foj,Shimizu:2016lbw}. When $Q_i$ strings are located at $x^{1,2,3,4}=0$, the Bianchi identity for the 2-form fields is modified as
\begin{align}
  dH^\alpha = X_4^\alpha + Q^\alpha \prod_{a=1}^4\delta(x^a)dx^a \ .
\end{align}
The shift in the right-hand side in the Bianchi identity applies to for the anomaly contribution from the Green-Schwarz term as
{\small
\begin{align}
I_8^{GS} \!=\! -\frac{1}{2}\Omega_{\alpha\beta}\!\left(\!X_4^\alpha\! +\!Q^\alpha \!\prod_{a=1}^4\!\delta(y^a)dy^a\!\right)\!\!\left(\!X_4^\beta\!+\! Q^\alpha\! \prod_{a=1}^4\!\delta(y^a)dy^a\!\right) \ .
\end{align}}
As a result, a non-trivial anomaly inflow is induced toward the string worldsheet. The anomaly inflow can be computed by integrating the 8-form anomaly polynomial over the 4 transverse directions to the strings. One computes
\begin{align}
  I_4^{\rm inflow} = -\Omega_{\alpha\beta}Q^\alpha\left(X_4^\alpha + \frac{1}{2}Q^\beta \chi_4(N_4)\right) \ .
\end{align}
This inflow must be cancelled by the anomalies arising from the worldsheet degrees of freedom on the 2d strings. Hence the anomaly polynomial of the 2d worldsheet SCFT must be
\begin{align}
  I_4 = -I_4^{\rm inflow} = \Omega_{\alpha\beta}Q^\alpha\left(X_4^\alpha + \frac{1}{2}Q^\beta \chi_4(N_4)\right) \ .
\end{align}
Here $\chi_4(N_4)$ is the Euler class of the $SO(4)=SU(2)_l\times SU(2)_R$ normal bundle for the transverse $\mathbb{R}^4$ directions and it can also be written as $\chi_4(N_4)=c_2(l)-c_2(R)$ in terms of the second Chern-classes $c_2(l)$ and $c_2(R)$ for $SU(2)_l$ and $SU(2)_R$.


\bibliography{ref}

\begin{thebibliography}{40}%
\makeatletter
\providecommand \@ifxundefined [1]{%
 \@ifx{#1\undefined}
}%
\providecommand \@ifnum [1]{%
 \ifnum #1\expandafter \@firstoftwo
 \else \expandafter \@secondoftwo
 \fi
}%
\providecommand \@ifx [1]{%
 \ifx #1\expandafter \@firstoftwo
 \else \expandafter \@secondoftwo
 \fi
}%
\providecommand \natexlab [1]{#1}%
\providecommand \enquote  [1]{``#1''}%
\providecommand \bibnamefont  [1]{#1}%
\providecommand \bibfnamefont [1]{#1}%
\providecommand \citenamefont [1]{#1}%
\providecommand \href@noop [0]{\@secondoftwo}%
\providecommand \href [0]{\begingroup \@sanitize@url \@href}%
\providecommand \@href[1]{\@@startlink{#1}\@@href}%
\providecommand \@@href[1]{\endgroup#1\@@endlink}%
\providecommand \@sanitize@url [0]{\catcode `\\12\catcode `\$12\catcode
  `\&12\catcode `\#12\catcode `\^12\catcode `\_12\catcode `\%12\relax}%
\providecommand \@@startlink[1]{}%
\providecommand \@@endlink[0]{}%
\providecommand \url  [0]{\begingroup\@sanitize@url \@url }%
\providecommand \@url [1]{\endgroup\@href {#1}{\urlprefix }}%
\providecommand \urlprefix  [0]{URL }%
\providecommand \Eprint [0]{\href }%
\providecommand \doibase [0]{http://dx.doi.org/}%
\providecommand \selectlanguage [0]{\@gobble}%
\providecommand \bibinfo  [0]{\@secondoftwo}%
\providecommand \bibfield  [0]{\@secondoftwo}%
\providecommand \translation [1]{[#1]}%
\providecommand \BibitemOpen [0]{}%
\providecommand \bibitemStop [0]{}%
\providecommand \bibitemNoStop [0]{.\EOS\space}%
\providecommand \EOS [0]{\spacefactor3000\relax}%
\providecommand \BibitemShut  [1]{\csname bibitem#1\endcsname}%
\let\auto@bib@innerbib\@empty
\bibitem [{\citenamefont {Brennan}\ \emph {et~al.}(2017)\citenamefont
  {Brennan}, \citenamefont {Carta},\ and\ \citenamefont
  {Vafa}}]{Brennan:2017rbf}%
  \BibitemOpen
  \bibfield  {author} {\bibinfo {author} {\bibfnamefont {T.~D.}\ \bibnamefont
  {Brennan}}, \bibinfo {author} {\bibfnamefont {F.}~\bibnamefont {Carta}}, \
  and\ \bibinfo {author} {\bibfnamefont {C.}~\bibnamefont {Vafa}},\ }\bibfield
  {booktitle} {\emph {\bibinfo {booktitle} {{Proceedings, Theoretical Advanced
  Study Institute in Elementary Particle Physics: Physics at the Fundamental
  Frontier (TASI 2017): Boulder, CO, USA, June 5-30, 2017}}},\ }\href {\doibase
  10.22323/1.305.0015} {\bibfield  {journal} {\bibinfo  {journal} {PoS}\
  }\textbf {\bibinfo {volume} {TASI2017}},\ \bibinfo {pages} {015} (\bibinfo
  {year} {2017})},\ \Eprint {http://arxiv.org/abs/1711.00864} {arXiv:1711.00864
  [hep-th]} \BibitemShut {NoStop}%
\bibitem [{\citenamefont {Palti}(2019)}]{Palti:2019pca}%
  \BibitemOpen
  \bibfield  {author} {\bibinfo {author} {\bibfnamefont {E.}~\bibnamefont
  {Palti}}\ }(\bibinfo {year} {2019})\ \Eprint
  {http://arxiv.org/abs/1903.06239} {arXiv:1903.06239 [hep-th]} \BibitemShut
  {NoStop}%
\bibitem [{\citenamefont {Banks}\ \emph {et~al.}(1997)\citenamefont {Banks},
  \citenamefont {Seiberg},\ and\ \citenamefont {Silverstein}}]{Banks:1997zs}%
  \BibitemOpen
  \bibfield  {author} {\bibinfo {author} {\bibfnamefont {T.}~\bibnamefont
  {Banks}}, \bibinfo {author} {\bibfnamefont {N.}~\bibnamefont {Seiberg}}, \
  and\ \bibinfo {author} {\bibfnamefont {E.}~\bibnamefont {Silverstein}},\
  }\href {\doibase 10.1016/S0370-2693(97)00366-3} {\bibfield  {journal}
  {\bibinfo  {journal} {Phys. Lett.}\ }\textbf {\bibinfo {volume} {B401}},\
  \bibinfo {pages} {30} (\bibinfo {year} {1997})},\ \Eprint
  {http://arxiv.org/abs/hep-th/9703052} {arXiv:hep-th/9703052 [hep-th]}
  \BibitemShut {NoStop}%
\bibitem [{\citenamefont {Taylor}(2011)}]{Taylor:2011wt}%
  \BibitemOpen
  \bibfield  {author} {\bibinfo {author} {\bibfnamefont {W.}~\bibnamefont
  {Taylor}},\ }\href@noop {} {\  (\bibinfo {year} {2011})},\ \Eprint
  {http://arxiv.org/abs/1104.2051} {arXiv:1104.2051 [hep-th]} \BibitemShut
  {NoStop}%
\bibitem [{\citenamefont {Green}\ \emph {et~al.}(1988)\citenamefont {Green},
  \citenamefont {Schwarz},\ and\ \citenamefont {Witten}}]{Green:1987mn}%
  \BibitemOpen
  \bibfield  {author} {\bibinfo {author} {\bibfnamefont {M.~B.}\ \bibnamefont
  {Green}}, \bibinfo {author} {\bibfnamefont {J.~H.}\ \bibnamefont {Schwarz}},
  \ and\ \bibinfo {author} {\bibfnamefont {E.}~\bibnamefont {Witten}},\ }\href
  {http://www.cambridge.org/us/academic/subjects/physics/theoretical-physics-and-mathematical-physics/superstring-theory-volume-2}
  {\emph {\bibinfo {title} {{SUPERSTRING THEORY. VOL. 2: LOOP AMPLITUDES,
  ANOMALIES AND PHENOMENOLOGY}}}}\ (\bibinfo {year} {1988})\BibitemShut
  {NoStop}%
\bibitem [{\citenamefont {Vafa}(2005)}]{Vafa:2005ui}%
  \BibitemOpen
  \bibfield  {author} {\bibinfo {author} {\bibfnamefont {C.}~\bibnamefont
  {Vafa}},\ }\href@noop {} {\  (\bibinfo {year} {2005})},\ \Eprint
  {http://arxiv.org/abs/hep-th/0509212} {arXiv:hep-th/0509212 [hep-th]}
  \BibitemShut {NoStop}%
\bibitem [{\citenamefont {Adams}\ \emph {et~al.}(2010)\citenamefont {Adams},
  \citenamefont {DeWolfe},\ and\ \citenamefont {Taylor}}]{Adams:2010zy}%
  \BibitemOpen
  \bibfield  {author} {\bibinfo {author} {\bibfnamefont {A.}~\bibnamefont
  {Adams}}, \bibinfo {author} {\bibfnamefont {O.}~\bibnamefont {DeWolfe}}, \
  and\ \bibinfo {author} {\bibfnamefont {W.}~\bibnamefont {Taylor}},\ }\href
  {\doibase 10.1103/PhysRevLett.105.071601} {\bibfield  {journal} {\bibinfo
  {journal} {Phys. Rev. Lett.}\ }\textbf {\bibinfo {volume} {105}},\ \bibinfo
  {pages} {071601} (\bibinfo {year} {2010})},\ \Eprint
  {http://arxiv.org/abs/1006.1352} {arXiv:1006.1352 [hep-th]} \BibitemShut
  {NoStop}%
\bibitem [{\citenamefont {Kumar}\ \emph
  {et~al.}(2010{\natexlab{a}})\citenamefont {Kumar}, \citenamefont {Morrison},\
  and\ \citenamefont {Taylor}}]{Kumar:2010ru}%
  \BibitemOpen
  \bibfield  {author} {\bibinfo {author} {\bibfnamefont {V.}~\bibnamefont
  {Kumar}}, \bibinfo {author} {\bibfnamefont {D.~R.}\ \bibnamefont {Morrison}},
  \ and\ \bibinfo {author} {\bibfnamefont {W.}~\bibnamefont {Taylor}},\ }\href
  {\doibase 10.1007/JHEP11(2010)118} {\bibfield  {journal} {\bibinfo  {journal}
  {JHEP}\ }\textbf {\bibinfo {volume} {11}},\ \bibinfo {pages} {118} (\bibinfo
  {year} {2010}{\natexlab{a}})},\ \Eprint {http://arxiv.org/abs/1008.1062}
  {arXiv:1008.1062 [hep-th]} \BibitemShut {NoStop}%
\bibitem [{\citenamefont {Polchinski}(2004)}]{Polchinski:2003bq}%
  \BibitemOpen
  \bibfield  {author} {\bibinfo {author} {\bibfnamefont {J.}~\bibnamefont
  {Polchinski}},\ }\bibfield  {booktitle} {\emph {\bibinfo {booktitle}
  {{Proceedings, Dirac Centennial Symposium, Tallahassee, USA, December 6-7,
  2002}}},\ }\href {\doibase 10.1142/S0217751X0401866X} {\bibfield  {journal}
  {\bibinfo  {journal} {Int. J. Mod. Phys.}\ }\textbf {\bibinfo {volume}
  {A19S1}},\ \bibinfo {pages} {145} (\bibinfo {year} {2004})},\ \bibinfo {note}
  {[,145(2003)]},\ \Eprint {http://arxiv.org/abs/hep-th/0304042}
  {arXiv:hep-th/0304042 [hep-th]} \BibitemShut {NoStop}%
\bibitem [{\citenamefont {Banks}\ and\ \citenamefont
  {Seiberg}(2011)}]{Banks:2010zn}%
  \BibitemOpen
  \bibfield  {author} {\bibinfo {author} {\bibfnamefont {T.}~\bibnamefont
  {Banks}}\ and\ \bibinfo {author} {\bibfnamefont {N.}~\bibnamefont
  {Seiberg}},\ }\href {\doibase 10.1103/PhysRevD.83.084019} {\bibfield
  {journal} {\bibinfo  {journal} {Phys. Rev.}\ }\textbf {\bibinfo {volume}
  {D83}},\ \bibinfo {pages} {084019} (\bibinfo {year} {2011})},\ \Eprint
  {http://arxiv.org/abs/1011.5120} {arXiv:1011.5120 [hep-th]} \BibitemShut
  {NoStop}%
\bibitem [{\citenamefont {Harlow}\ and\ \citenamefont
  {Ooguri}(2018)}]{Harlow:2018tng}%
  \BibitemOpen
  \bibfield  {author} {\bibinfo {author} {\bibfnamefont {D.}~\bibnamefont
  {Harlow}}\ and\ \bibinfo {author} {\bibfnamefont {H.}~\bibnamefont
  {Ooguri}},\ }\href@noop {} {\  (\bibinfo {year} {2018})},\ \Eprint
  {http://arxiv.org/abs/1810.05338} {arXiv:1810.05338 [hep-th]} \BibitemShut
  {NoStop}%
\bibitem [{\citenamefont {Green}\ and\ \citenamefont
  {Schwarz}(1984)}]{Green:1984sg}%
  \BibitemOpen
  \bibfield  {author} {\bibinfo {author} {\bibfnamefont {M.~B.}\ \bibnamefont
  {Green}}\ and\ \bibinfo {author} {\bibfnamefont {J.~H.}\ \bibnamefont
  {Schwarz}},\ }\href {\doibase 10.1016/0370-2693(84)91565-X} {\bibfield
  {journal} {\bibinfo  {journal} {Phys. Lett.}\ }\textbf {\bibinfo {volume}
  {149B}},\ \bibinfo {pages} {117} (\bibinfo {year} {1984})}\BibitemShut
  {NoStop}%
\bibitem [{\citenamefont {Callan}\ and\ \citenamefont
  {Harvey}(1985)}]{Callan:1984sa}%
  \BibitemOpen
  \bibfield  {author} {\bibinfo {author} {\bibfnamefont {C.~G.}\ \bibnamefont
  {Callan}, \bibfnamefont {Jr.}}\ and\ \bibinfo {author} {\bibfnamefont
  {J.~A.}\ \bibnamefont {Harvey}},\ }\href {\doibase
  10.1016/0550-3213(85)90489-4} {\bibfield  {journal} {\bibinfo  {journal}
  {Nucl. Phys.}\ }\textbf {\bibinfo {volume} {B250}},\ \bibinfo {pages} {427}
  (\bibinfo {year} {1985})}\BibitemShut {NoStop}%
\bibitem [{\citenamefont {Blum}\ and\ \citenamefont
  {Harvey}(1994)}]{Blum:1993yd}%
  \BibitemOpen
  \bibfield  {author} {\bibinfo {author} {\bibfnamefont {J.~D.}\ \bibnamefont
  {Blum}}\ and\ \bibinfo {author} {\bibfnamefont {J.~A.}\ \bibnamefont
  {Harvey}},\ }\href {\doibase 10.1016/0550-3213(94)90580-0} {\bibfield
  {journal} {\bibinfo  {journal} {Nucl. Phys.}\ }\textbf {\bibinfo {volume}
  {B416}},\ \bibinfo {pages} {119} (\bibinfo {year} {1994})},\ \Eprint
  {http://arxiv.org/abs/hep-th/9310035} {arXiv:hep-th/9310035 [hep-th]}
  \BibitemShut {NoStop}%
\bibitem [{\citenamefont {Freed}\ \emph {et~al.}(1998)\citenamefont {Freed},
  \citenamefont {Harvey}, \citenamefont {Minasian},\ and\ \citenamefont
  {Moore}}]{Freed:1998tg}%
  \BibitemOpen
  \bibfield  {author} {\bibinfo {author} {\bibfnamefont {D.}~\bibnamefont
  {Freed}}, \bibinfo {author} {\bibfnamefont {J.~A.}\ \bibnamefont {Harvey}},
  \bibinfo {author} {\bibfnamefont {R.}~\bibnamefont {Minasian}}, \ and\
  \bibinfo {author} {\bibfnamefont {G.~W.}\ \bibnamefont {Moore}},\ }\href
  {\doibase 10.4310/ATMP.1998.v2.n3.a8} {\bibfield  {journal} {\bibinfo
  {journal} {Adv. Theor. Math. Phys.}\ }\textbf {\bibinfo {volume} {2}},\
  \bibinfo {pages} {601} (\bibinfo {year} {1998})},\ \Eprint
  {http://arxiv.org/abs/hep-th/9803205} {arXiv:hep-th/9803205 [hep-th]}
  \BibitemShut {NoStop}%
\bibitem [{\citenamefont {Bergshoeff}\ \emph {et~al.}(1982)\citenamefont
  {Bergshoeff}, \citenamefont {de~Roo}, \citenamefont {de~Wit},\ and\
  \citenamefont {van Nieuwenhuizen}}]{Bergshoeff:1981um}%
  \BibitemOpen
  \bibfield  {author} {\bibinfo {author} {\bibfnamefont {E.}~\bibnamefont
  {Bergshoeff}}, \bibinfo {author} {\bibfnamefont {M.}~\bibnamefont {de~Roo}},
  \bibinfo {author} {\bibfnamefont {B.}~\bibnamefont {de~Wit}}, \ and\ \bibinfo
  {author} {\bibfnamefont {P.}~\bibnamefont {van Nieuwenhuizen}},\ }\href
  {\doibase 10.1016/0550-3213(82)90050-5} {\bibfield  {journal} {\bibinfo
  {journal} {Nucl. Phys.}\ }\textbf {\bibinfo {volume} {B195}},\ \bibinfo
  {pages} {97} (\bibinfo {year} {1982})}\BibitemShut {NoStop}%
\bibitem [{\citenamefont {Chapline}\ and\ \citenamefont
  {Manton}(1983)}]{Chapline:1982ww}%
  \BibitemOpen
  \bibfield  {author} {\bibinfo {author} {\bibfnamefont {G.~F.}\ \bibnamefont
  {Chapline}}\ and\ \bibinfo {author} {\bibfnamefont {N.~S.}\ \bibnamefont
  {Manton}},\ }\href {\doibase 10.1016/0370-2693(83)90633-0} {\bibfield
  {journal} {\bibinfo  {journal} {Phys. Lett.}\ }\textbf {\bibinfo {volume}
  {B120}},\ \bibinfo {pages} {105} (\bibinfo {year} {1983})},\ \bibinfo {note}
  {[,105(1982)]}\BibitemShut {NoStop}%
\bibitem [{\citenamefont {Di~Francesco}\ \emph {et~al.}(1997)\citenamefont
  {Di~Francesco}, \citenamefont {Mathieu},\ and\ \citenamefont
  {Senechal}}]{DiFrancesco:1997nk}%
  \BibitemOpen
  \bibfield  {author} {\bibinfo {author} {\bibfnamefont {P.}~\bibnamefont
  {Di~Francesco}}, \bibinfo {author} {\bibfnamefont {P.}~\bibnamefont
  {Mathieu}}, \ and\ \bibinfo {author} {\bibfnamefont {D.}~\bibnamefont
  {Senechal}},\ }\href {\doibase 10.1007/978-1-4612-2256-9} {\emph {\bibinfo
  {title} {{Conformal Field Theory}}}},\ Graduate Texts in Contemporary
  Physics\ (\bibinfo  {publisher} {Springer-Verlag},\ \bibinfo {address} {New
  York},\ \bibinfo {year} {1997})\BibitemShut {NoStop}%
\bibitem [{\citenamefont {Sagnotti}(1992)}]{Sagnotti:1992qw}%
  \BibitemOpen
  \bibfield  {author} {\bibinfo {author} {\bibfnamefont {A.}~\bibnamefont
  {Sagnotti}},\ }\href {\doibase 10.1016/0370-2693(92)90682-T} {\bibfield
  {journal} {\bibinfo  {journal} {Phys. Lett.}\ }\textbf {\bibinfo {volume}
  {B294}},\ \bibinfo {pages} {196} (\bibinfo {year} {1992})},\ \Eprint
  {http://arxiv.org/abs/hep-th/9210127} {arXiv:hep-th/9210127 [hep-th]}
  \BibitemShut {NoStop}%
\bibitem [{\citenamefont {Schwarz}(1996)}]{Schwarz:1995zw}%
  \BibitemOpen
  \bibfield  {author} {\bibinfo {author} {\bibfnamefont {J.~H.}\ \bibnamefont
  {Schwarz}},\ }\href {\doibase 10.1016/0370-2693(95)01610-4} {\bibfield
  {journal} {\bibinfo  {journal} {Phys. Lett.}\ }\textbf {\bibinfo {volume}
  {B371}},\ \bibinfo {pages} {223} (\bibinfo {year} {1996})},\ \Eprint
  {http://arxiv.org/abs/hep-th/9512053} {arXiv:hep-th/9512053 [hep-th]}
  \BibitemShut {NoStop}%
\bibitem [{\citenamefont {Kumar}\ and\ \citenamefont
  {Taylor}(2011)}]{Kumar:2009us}%
  \BibitemOpen
  \bibfield  {author} {\bibinfo {author} {\bibfnamefont {V.}~\bibnamefont
  {Kumar}}\ and\ \bibinfo {author} {\bibfnamefont {W.}~\bibnamefont {Taylor}},\
  }\href {\doibase 10.4310/ATMP.2011.v15.n2.a3} {\bibfield  {journal} {\bibinfo
   {journal} {Adv. Theor. Math. Phys.}\ }\textbf {\bibinfo {volume} {15}},\
  \bibinfo {pages} {325} (\bibinfo {year} {2011})},\ \Eprint
  {http://arxiv.org/abs/0906.0987} {arXiv:0906.0987 [hep-th]} \BibitemShut
  {NoStop}%
\bibitem [{\citenamefont {Kumar}\ and\ \citenamefont
  {Taylor}(2009)}]{Kumar:2009ae}%
  \BibitemOpen
  \bibfield  {author} {\bibinfo {author} {\bibfnamefont {V.}~\bibnamefont
  {Kumar}}\ and\ \bibinfo {author} {\bibfnamefont {W.}~\bibnamefont {Taylor}},\
  }\href {\doibase 10.1088/1126-6708/2009/12/050} {\bibfield  {journal}
  {\bibinfo  {journal} {JHEP}\ }\textbf {\bibinfo {volume} {12}},\ \bibinfo
  {pages} {050} (\bibinfo {year} {2009})},\ \Eprint
  {http://arxiv.org/abs/0910.1586} {arXiv:0910.1586 [hep-th]} \BibitemShut
  {NoStop}%
\bibitem [{\citenamefont {Kumar}\ \emph {et~al.}(2011)\citenamefont {Kumar},
  \citenamefont {Park},\ and\ \citenamefont {Taylor}}]{Kumar:2010am}%
  \BibitemOpen
  \bibfield  {author} {\bibinfo {author} {\bibfnamefont {V.}~\bibnamefont
  {Kumar}}, \bibinfo {author} {\bibfnamefont {D.~S.}\ \bibnamefont {Park}}, \
  and\ \bibinfo {author} {\bibfnamefont {W.}~\bibnamefont {Taylor}},\ }\href
  {\doibase 10.1007/JHEP04(2011)080} {\bibfield  {journal} {\bibinfo  {journal}
  {JHEP}\ }\textbf {\bibinfo {volume} {04}},\ \bibinfo {pages} {080} (\bibinfo
  {year} {2011})},\ \Eprint {http://arxiv.org/abs/1011.0726} {arXiv:1011.0726
  [hep-th]} \BibitemShut {NoStop}%
\bibitem [{\citenamefont {Johnson}\ and\ \citenamefont
  {Taylor}(2016)}]{Johnson:2016qar}%
  \BibitemOpen
  \bibfield  {author} {\bibinfo {author} {\bibfnamefont {S.~B.}\ \bibnamefont
  {Johnson}}\ and\ \bibinfo {author} {\bibfnamefont {W.}~\bibnamefont
  {Taylor}},\ }\href {\doibase 10.1002/prop.201600074} {\bibfield  {journal}
  {\bibinfo  {journal} {Fortsch. Phys.}\ }\textbf {\bibinfo {volume} {64}},\
  \bibinfo {pages} {581} (\bibinfo {year} {2016})},\ \Eprint
  {http://arxiv.org/abs/1605.08052} {arXiv:1605.08052 [hep-th]} \BibitemShut
  {NoStop}%
\bibitem [{\citenamefont {Taylor}\ and\ \citenamefont
  {Turner}(2018)}]{Taylor:2018khc}%
  \BibitemOpen
  \bibfield  {author} {\bibinfo {author} {\bibfnamefont {W.}~\bibnamefont
  {Taylor}}\ and\ \bibinfo {author} {\bibfnamefont {A.~P.}\ \bibnamefont
  {Turner}},\ }\href {\doibase 10.1007/JHEP06(2018)010} {\bibfield  {journal}
  {\bibinfo  {journal} {JHEP}\ }\textbf {\bibinfo {volume} {06}},\ \bibinfo
  {pages} {010} (\bibinfo {year} {2018})},\ \Eprint
  {http://arxiv.org/abs/1803.04447} {arXiv:1803.04447 [hep-th]} \BibitemShut
  {NoStop}%
\bibitem [{\citenamefont {Raghuram}\ and\ \citenamefont
  {Taylor}(2018)}]{Raghuram:2018hjn}%
  \BibitemOpen
  \bibfield  {author} {\bibinfo {author} {\bibfnamefont {N.}~\bibnamefont
  {Raghuram}}\ and\ \bibinfo {author} {\bibfnamefont {W.}~\bibnamefont
  {Taylor}},\ }\href {\doibase 10.1007/JHEP10(2018)182} {\bibfield  {journal}
  {\bibinfo  {journal} {JHEP}\ }\textbf {\bibinfo {volume} {10}},\ \bibinfo
  {pages} {182} (\bibinfo {year} {2018})},\ \Eprint
  {http://arxiv.org/abs/1809.01666} {arXiv:1809.01666 [hep-th]} \BibitemShut
  {NoStop}%
\bibitem [{\citenamefont {Taylor}\ and\ \citenamefont
  {Turner}(2019)}]{Taylor:2019ots}%
  \BibitemOpen
  \bibfield  {author} {\bibinfo {author} {\bibfnamefont {W.}~\bibnamefont
  {Taylor}}\ and\ \bibinfo {author} {\bibfnamefont {A.~P.}\ \bibnamefont
  {Turner}},\ }\href@noop {} {\  (\bibinfo {year} {2019})},\ \Eprint
  {http://arxiv.org/abs/1901.02012} {arXiv:1901.02012 [hep-th]} \BibitemShut
  {NoStop}%
\bibitem [{\citenamefont {Kim}\ \emph {et~al.}(2016)\citenamefont {Kim},
  \citenamefont {Kim},\ and\ \citenamefont {Park}}]{Kim:2016foj}%
  \BibitemOpen
  \bibfield  {author} {\bibinfo {author} {\bibfnamefont {H.-C.}\ \bibnamefont
  {Kim}}, \bibinfo {author} {\bibfnamefont {S.}~\bibnamefont {Kim}}, \ and\
  \bibinfo {author} {\bibfnamefont {J.}~\bibnamefont {Park}},\ }\href@noop {}
  {\  (\bibinfo {year} {2016})},\ \Eprint {http://arxiv.org/abs/1608.03919}
  {arXiv:1608.03919 [hep-th]} \BibitemShut {NoStop}%
\bibitem [{\citenamefont {Shimizu}\ and\ \citenamefont
  {Tachikawa}(2016)}]{Shimizu:2016lbw}%
  \BibitemOpen
  \bibfield  {author} {\bibinfo {author} {\bibfnamefont {H.}~\bibnamefont
  {Shimizu}}\ and\ \bibinfo {author} {\bibfnamefont {Y.}~\bibnamefont
  {Tachikawa}},\ }\href {\doibase 10.1007/JHEP11(2016)165} {\bibfield
  {journal} {\bibinfo  {journal} {JHEP}\ }\textbf {\bibinfo {volume} {11}},\
  \bibinfo {pages} {165} (\bibinfo {year} {2016})},\ \Eprint
  {http://arxiv.org/abs/1608.05894} {arXiv:1608.05894 [hep-th]} \BibitemShut
  {NoStop}%
\bibitem [{\citenamefont {Hayashi}\ \emph {et~al.}(2019)\citenamefont
  {Hayashi}, \citenamefont {Jefferson}, \citenamefont {Kim}, \citenamefont
  {Ohmori},\ and\ \citenamefont {Vafa}}]{Hayashi:2019fsa}%
  \BibitemOpen
  \bibfield  {author} {\bibinfo {author} {\bibfnamefont {H.}~\bibnamefont
  {Hayashi}}, \bibinfo {author} {\bibfnamefont {P.}~\bibnamefont {Jefferson}},
  \bibinfo {author} {\bibfnamefont {H.-C.}\ \bibnamefont {Kim}}, \bibinfo
  {author} {\bibfnamefont {K.}~\bibnamefont {Ohmori}}, \ and\ \bibinfo {author}
  {\bibfnamefont {C.}~\bibnamefont {Vafa}},\ }\href@noop {} {\  (\bibinfo
  {year} {2019})},\ \Eprint {http://arxiv.org/abs/1905.00116} {arXiv:1905.00116
  [hep-th]} \BibitemShut {NoStop}%
\bibitem [{\citenamefont {Haghighat}\ \emph {et~al.}(2016)\citenamefont
  {Haghighat}, \citenamefont {Murthy}, \citenamefont {Vafa},\ and\
  \citenamefont {Vandoren}}]{Haghighat:2015ega}%
  \BibitemOpen
  \bibfield  {author} {\bibinfo {author} {\bibfnamefont {B.}~\bibnamefont
  {Haghighat}}, \bibinfo {author} {\bibfnamefont {S.}~\bibnamefont {Murthy}},
  \bibinfo {author} {\bibfnamefont {C.}~\bibnamefont {Vafa}}, \ and\ \bibinfo
  {author} {\bibfnamefont {S.}~\bibnamefont {Vandoren}},\ }\href {\doibase
  10.1007/JHEP01(2016)009} {\bibfield  {journal} {\bibinfo  {journal} {JHEP}\
  }\textbf {\bibinfo {volume} {01}},\ \bibinfo {pages} {009} (\bibinfo {year}
  {2016})},\ \Eprint {http://arxiv.org/abs/1509.00455} {arXiv:1509.00455
  [hep-th]} \BibitemShut {NoStop}%
\bibitem [{Note1()}]{Note1}%
  \BibitemOpen
  \bibinfo {note} {The coefficients of the Gauss-Bonnet term, the
  Riemann-squared term, and the Weyl-squared term are all equal. This can be
  seen using the equations of motion to rewrite the operators involving $R$ and
  $R_{\mu \nu }$.}\BibitemShut {Stop}%
\bibitem [{\citenamefont {Cheung}\ and\ \citenamefont
  {Remmen}(2017)}]{Cheung:2016wjt}%
  \BibitemOpen
  \bibfield  {author} {\bibinfo {author} {\bibfnamefont {C.}~\bibnamefont
  {Cheung}}\ and\ \bibinfo {author} {\bibfnamefont {G.~N.}\ \bibnamefont
  {Remmen}},\ }\href {\doibase 10.1103/PhysRevLett.118.051601} {\bibfield
  {journal} {\bibinfo  {journal} {Phys. Rev. Lett.}\ }\textbf {\bibinfo
  {volume} {118}},\ \bibinfo {pages} {051601} (\bibinfo {year} {2017})},\
  \Eprint {http://arxiv.org/abs/1608.02942} {arXiv:1608.02942 [hep-th]}
  \BibitemShut {NoStop}%
\bibitem [{\citenamefont {Hamada}\ \emph {et~al.}(2018)\citenamefont {Hamada},
  \citenamefont {Noumi},\ and\ \citenamefont {Shiu}}]{Hamada:2018dde}%
  \BibitemOpen
  \bibfield  {author} {\bibinfo {author} {\bibfnamefont {Y.}~\bibnamefont
  {Hamada}}, \bibinfo {author} {\bibfnamefont {T.}~\bibnamefont {Noumi}}, \
  and\ \bibinfo {author} {\bibfnamefont {G.}~\bibnamefont {Shiu}},\ }\href@noop
  {} {\  (\bibinfo {year} {2018})},\ \Eprint {http://arxiv.org/abs/1810.03637}
  {arXiv:1810.03637 [hep-th]} \BibitemShut {NoStop}%
\bibitem [{Note2()}]{Note2}%
  \BibitemOpen
  \bibinfo {note} {This last condition in F-theory setup translates to the
  condition that $J\cdot K <0$, signifying that the base of F-theory
  compactification is positively curved, which is necessary for solving
  Einstein's equation when $\tau $ varies over the base.}\BibitemShut {Stop}%
\bibitem [{\citenamefont {Dabholkar}\ and\ \citenamefont
  {Park}(1996{\natexlab{a}})}]{Dabholkar:1996zi}%
  \BibitemOpen
  \bibfield  {author} {\bibinfo {author} {\bibfnamefont {A.}~\bibnamefont
  {Dabholkar}}\ and\ \bibinfo {author} {\bibfnamefont {J.}~\bibnamefont
  {Park}},\ }\href {\doibase 10.1016/0550-3213(96)00199-X} {\bibfield
  {journal} {\bibinfo  {journal} {Nucl. Phys.}\ }\textbf {\bibinfo {volume}
  {B472}},\ \bibinfo {pages} {207} (\bibinfo {year} {1996}{\natexlab{a}})},\
  \Eprint {http://arxiv.org/abs/hep-th/9602030} {arXiv:hep-th/9602030 [hep-th]}
  \BibitemShut {NoStop}%
\bibitem [{\citenamefont {Dabholkar}\ and\ \citenamefont
  {Park}(1996{\natexlab{b}})}]{Dabholkar:1996pc}%
  \BibitemOpen
  \bibfield  {author} {\bibinfo {author} {\bibfnamefont {A.}~\bibnamefont
  {Dabholkar}}\ and\ \bibinfo {author} {\bibfnamefont {J.}~\bibnamefont
  {Park}},\ }\href {\doibase 10.1016/0550-3213(96)00395-1} {\bibfield
  {journal} {\bibinfo  {journal} {Nucl. Phys.}\ }\textbf {\bibinfo {volume}
  {B477}},\ \bibinfo {pages} {701} (\bibinfo {year} {1996}{\natexlab{b}})},\
  \Eprint {http://arxiv.org/abs/hep-th/9604178} {arXiv:hep-th/9604178 [hep-th]}
  \BibitemShut {NoStop}%
\bibitem [{\citenamefont {Kumar}\ \emph
  {et~al.}(2010{\natexlab{b}})\citenamefont {Kumar}, \citenamefont {Morrison},\
  and\ \citenamefont {Taylor}}]{Kumar:2009ac}%
  \BibitemOpen
  \bibfield  {author} {\bibinfo {author} {\bibfnamefont {V.}~\bibnamefont
  {Kumar}}, \bibinfo {author} {\bibfnamefont {D.~R.}\ \bibnamefont {Morrison}},
  \ and\ \bibinfo {author} {\bibfnamefont {W.}~\bibnamefont {Taylor}},\ }\href
  {\doibase 10.1007/JHEP02(2010)099} {\bibfield  {journal} {\bibinfo  {journal}
  {JHEP}\ }\textbf {\bibinfo {volume} {02}},\ \bibinfo {pages} {099} (\bibinfo
  {year} {2010}{\natexlab{b}})},\ \Eprint {http://arxiv.org/abs/0911.3393}
  {arXiv:0911.3393 [hep-th]} \BibitemShut {NoStop}%
\bibitem [{\citenamefont {Seiberg}\ and\ \citenamefont
  {Witten}(1996)}]{Seiberg:1996vs}%
  \BibitemOpen
  \bibfield  {author} {\bibinfo {author} {\bibfnamefont {N.}~\bibnamefont
  {Seiberg}}\ and\ \bibinfo {author} {\bibfnamefont {E.}~\bibnamefont
  {Witten}},\ }\href {\doibase 10.1016/0550-3213(96)00189-7} {\bibfield
  {journal} {\bibinfo  {journal} {Nucl. Phys.}\ }\textbf {\bibinfo {volume}
  {B471}},\ \bibinfo {pages} {121} (\bibinfo {year} {1996})},\ \Eprint
  {http://arxiv.org/abs/hep-th/9603003} {arXiv:hep-th/9603003 [hep-th]}
  \BibitemShut {NoStop}%
\bibitem [{\citenamefont {Bilal}(2008)}]{Bilal:2008qx}%
  \BibitemOpen
  \bibfield  {author} {\bibinfo {author} {\bibfnamefont {A.}~\bibnamefont
  {Bilal}},\ }\href@noop {} {\  (\bibinfo {year} {2008})},\ \Eprint
  {http://arxiv.org/abs/0802.0634} {arXiv:0802.0634 [hep-th]} \BibitemShut
  {NoStop}%
\end{thebibliography}%
\end{document}